\documentclass[journal=jacsat,manuscript=article]{achemso}

\usepackage[version=3]{mhchem} 
\usepackage[T1]{fontenc}       

\usepackage{dcolumn}
\usepackage{bm}
\usepackage{amsmath}
\usepackage{amsfonts}
\usepackage{amssymb}
\usepackage{graphicx}
\usepackage{tabularx}
\usepackage{subfigure}
\usepackage{color}
\usepackage{gensymb}
\usepackage{physics}
\usepackage{hyperref}

\author{Sobhit Singh}
\affiliation{Department of Physics and Astronomy, West Virginia University, Morgantown, WV-26505-6315, USA}
\email{smsingh@mix.wvu.edu}

\author{Aldo H. Romero}
\affiliation{Department of Physics and Astronomy, West Virginia University, Morgantown, WV-26505-6315, USA}
\email{Aldo.Romero@mail.wvu.edu}

\title[An \textsf{achemso} demo]
  {Giant tunable Rashba spin splitting in two-dimensional BiSb monolayer and BiSb/AlN heterostructures}


\begin{document}






\begin{abstract}
Search of novel two-dimensional giant Rashba semiconductors is a crucial step in the development of the forthcoming nano-spintronics technology. Using first-principle calculations, we study a stable two-dimensional crystal phase of BiSb having buckled honeycomb lattice geometry, which is yet unexplored. The phonon, room temperature molecular dynamics and elastic constant calculations verify the dynamical and mechanical stability of the monolayer at 0~K and at room temperature. The calculated electronic bandstructure reveals the direct bandgap semiconducting nature of BiSb monolayer with presence of highly mobile two-dimensional electron gas (2DEG) near Fermi-level. Inclusion of spin-orbit coupling (SOC) yields the giant Rashba spin-splitting of 2DEG near Fermi-level. The calculated Rashba energy and Rashba splitting constant are 13 meV and 2.3 eV\AA, respectively. The strength of the Rashba splitting is amongst the largest yet known 2D Rashba semiconductors. We demonstrate that the strength of the Rashba spin-splitting can be significantly tuned by applying in-plane bi-axial strain on the BiSb monolayer. Presence of the giant Rashba spin-splitting together with the large electronic bandgap (1.6 eV) makes this system of peculiar interest for optoelectronics applications. Furthermore, we study the electronic properties of BiSb/AlN heterostructures having a lattice mismatch of 1.3\% at the interface. Our results suggest that BiSb monolayer and heterostructure systems could be potentially used to develop highly efficient spin field-effect transistors, optoelectronics and nano-spintronics devices. Thus, this comprehensive study of two-dimensional BiSb systems can expand the range of possible applications in the future spintronics technology.  
\end{abstract}

\section{Introduction}
Discovery of new materials having large spin-orbit coupling (SOC) is very crucial in the rapidly burgeoning field of spintronics. In spintronics, we exploit the spin-orbit interaction present in materials to tune their electronic properties. The spin-orbit interaction originates due to the relativistic motion of electrons and acts as an effective built-in magnetic field in non-magnetic materials, which functions similar to the external magnetic field in the celebrated quantum Hall effect. \cite{quantumhalleffect1980, quantumhalleffect1982} Electrons with opposite spins feel opposite magnetic field in their rest frame and this field couples to their magnetic moment. Nevertheless, the net effects do not cancel out but yield a new quantum phenomenon called Rashba effect.\cite{rashba1960} The Rashba effect appears in systems with broken inversion-symmetry and causes lift in the spin degeneracy of electronic bands. Initially, the Rashba effect was believed to arise at the surfaces and interfaces due to the asymmetry of the confinement potential. However, recent works reveal that the Rashba effect can also be realized in bulk semiconductors. \cite{IshizakaNature2011, BahramyPRB2011, silvia_bulkgete_2013, SnTe_PRBsilvia2014, XiangPRBbitecl_2015, NarayanPRB2015, PhysRevB_silveira2016} 

The strength of the Rashba effect can be quantized by three key parameters: Rashba energy ($E_R$), Rashba momentum ($k_o$) and Rashba constant ($\alpha_R$). Materials with large Rashba energy and Rashba constant provide us more opportunities to tune the spintronics properties. Experiments report the presence of 2D Rashba spin-splitting in InAlAs/InGaAs\cite{Nitta_PRL1997} and LaAlO$_3$/SrTiO$_3$\cite{CavigliaPRL2010, ZhongPRB2013} interfaces. However, the magnitude of the Rashba constant ($\alpha_R$) is not large in the mentioned interfaces ($\alpha_R$ = $0.07$ eV\AA~for InAlAs/InGaAs \cite{Nitta_PRL1997} and $\alpha_R$ = $0.01-0.05$ eV\AA~for LaAlO$_3$/SrTiO$_3$ oxide interface\cite{CavigliaPRL2010, ZhongPRB2013}). Currently, surfaces of heavy metals such as Au, Bi, Ir and BiAg(111) alloys are known to exhibit large Rashba spin-splitting. The magnitude of $\alpha_R$ in Au(111)\cite{LaShell_AuPRL}, Bi(111)\cite{Koroteev_PRL2004}, Ir(111)\cite{Varykhalov_Ir_PRL2012} and BiAg(111)\cite{Ast_BiSbPRL2007} surfaces is 0.33, 0.55, 1.3 and 3.05 eV\AA, respectively. Even though these surfaces inherit large Rashba spin-splitting, yet they cannot be used in many spintronics device applications due to the presence of the (semi)metallic surface states. In particular, there is still no stable 2D semiconductor having large Rashba effect which is suitable for spin field-effect transistor applications. 

In addition to the spintronics applications, the 2D semiconductors inheriting large Rashba effect are strongly desired to hunt Majorana fermions. A 2D Rashba semiconductor interfaced with an $s$-wave superconductor under broken time-reversal symmetry can be used to build topological heterostructures that can support the long-sought but not yet detected Majorana fermions.\cite{Sau_PRL2010, Sau_PRB2010, Jason_PRB2010} Majorana fermions manifest themselves as a zero-bias conductance peak.\cite{Das_Nature2012} If experimentally realized, Majorana fermions may pave the way for realization of fault-tolerant topological quantum computation. \cite{Kitaev2003, JCTeo_PRL2010, JDSau_PRB2011} Therefore, there is a huge demand to search for new stable materials inheriting large controllable Rashba effect in two-dimensions.

In present work, we study the electronic properties of BiSb monolayer by means of first-principle calculations. Our calculations reveal that BiSb monolayer forms a stable free-standing 2D crystal and exhibits giant controllable Rashba effect. The observed $E_R$ and $\alpha_R$ are 137 meV and 2.3 eV\AA, respectively, which is amongst the largest yet known Rashba spin-splitting parameters in two-dimensional materials.\cite{Nitta_PRL1997, CavigliaPRL2010, ZhongPRB2013, LaShell_AuPRL, Koroteev_PRL2004, Varykhalov_Ir_PRL2012, Ast_BiSbPRL2007, Liu_NanoLetter2013} We demonstrate that the Rashba effect in BiSb monolayer can be efficiently tuned by applying bi-axial strain. The calculated energy bandgap is 1.6 eV, which lies in the visible range. Presence of the giant Rashba spin-splitting together with the large electronic bandgap makes this system of peculiar interest for optoelectronics applications. We further investigate the electronic properties of BiSb monolayer placed in contact with an AlN substrate. We construct BiSb/AlN heterostructures with two possible layer terminations (Bi and Sb) at the interface. Both BiSb/AlN van der Waal (vdW) heterostructures are found to exhibit direct bandgap semiconducting nature with presence of light mobile two-dimensional electron gas (2DEG) at the interface. Remarkably, we notice signatures of the giant Rashba spin-splitting of 2DEG near the Fermi-level. Our results suggest that BiSb monolayer and BiSb/AlN heterostructure systems could be potentially used to develop highly efficient spin field-effect transistors, optoelectronics and nano-spintronics devices.

\section{Computational Details}
Density Functional Theory (DFT) based first-principle calculations were carried out using the projector augmented-wave (PAW) method as implemented in the {\sc VASP} code \cite{Kresse1996, Kresse1999}. We used the PBE exchange-correlation functional as parametrized by Perdew-Burke-Ernzerhof \cite{Perdew1996}. The SOC was employed by a second-variation method implemented in the {\sc VASP} code. We considered fifteen valence electrons of Bi ($5d^{10}6s^{2}6p^{3}$) and five valence electrons of Sb ($5s^{2}5p^{3}$) in the PAW pseudo-potential. The lattice parameters of the monolayer were optimized until the Hellmann-Feynman residual forces were less than $10^{-4}$ eV/{\AA} per atom. For convergence of the electronic self-consistent calculations, a total energy difference criterion was defined as $10^{-8}$ eV. We used $650$ eV as kinetic energy cutoff of the plane wave basis set and a $\Gamma$-type $10\cross10\cross1$ $ k$-point mesh was employed to optimize the lattice parameters and the self-energy. The phonon calculations were performed for a $4\cross4\cross1$ supercell using the Density Functional Perturbation Theory (DFPT) approach as implemented in {\sc VASP} code. The PHONOPY code \cite{Togo2008} was used for the phonons post-processing. SOC was included in the phonon calculations as well as in the optimization of the primitive cell. To verify the stability of the BiSb monolayer at room temperature, we performed room temperature molecular dynamics (MD) simulations for more than 6000 fs with a time step of 1 fs. In MD simulations, we employed a supercell of size $4\cross4\cross1$ to guarantee the decay of interatomic force constants within the supercell dimensions. To investigate the effect of in-plane bi-axial strain, we varied the $a$ and $b$ lattice vectors from $-8\%$ (compression) to $+8\%$ (expansion) while performing relaxation of the inner-coordinates. The screened hybrid Heyd-Scuseria-Ernzerhof (HSE06) functional\cite{HSE_2003, Heyd2004} was used to get a better estimation of the electronic bandgap of BiSb monolayer. The Bi- and Sb-terminated BiSb/AlN heterostructures were modeled by using supercells of size ($3\cross3\cross1$)/($4\cross4\cross1$). This combination yields a lattice mismatch of 1.3\%. The vdW interaction\cite{TS_PRL2009, Bu_TSvdW2013} together with SOC as implemented in the {\sc VASP} code were included in the structural optimization as well as in the electronic structure calculations reported for BiSb/AlN heterostructures. The heterostructures were optimized until the total residual forces on each atom were less than 0.001 eV/\AA. A $\Gamma$ k-point mesh of size $8\cross8\cross1$ was used to sample the irreducible Brillouin zone of heterostructures. A vacuum of thickness larger than 15 \AA~was added along $c$-axis to avoid any interaction between two periodic BiSb and BiSb/AlN geometries. The PyProcar code was used to calculate the constant energy contour plots of the spin-texture.\cite{PyProcar} 

\section{Results and discussion}

\subsection{Crystal structure and the stability of BiSb monolayer}
The bulk BiSb compound is known to exhibit a rhombohedral structure with $R3m$ symmetry in its ground state.\cite{singh2016PCCP, sobhit2016PRB} In this geometry, the Bi and Sb atoms are stacked along the (111) direction of the rhombohedral primitive cell. Two alternative Bi and Sb layers strongly interact covalently and form a BiSb bilayer, whereas, two adjacent BiSb bilayers weekly interact due to the weak vdW interaction. Therefore, it is possible to fabricate a stable two-dimensional BiSb monolayer by advanced exfoliation, vapor deposition, or molecular-beam epitaxy techniques. A detailed study of the structural, vibrational, electronic, thermoelectric and topological behaviour of bulk BiSb can be found in Ref \cite{singh2016PCCP, sobhit2016PRB}. Bulk BiSb exhibits a giant Rashba effect with presence of the pressure-driven Weyl semimetallic phase.\cite{sobhit2016PRB} This is one of the first predicted ferroelectric Weyl semimetals which demonstrates tunability of Weyl charges. Such unique and intriguing topological features of bulk BiSb stimulate research in the two-dimensional BiSb. 

\begin{figure}[hb!]
 \centering
 \includegraphics[width=17.0cm, keepaspectratio=true]{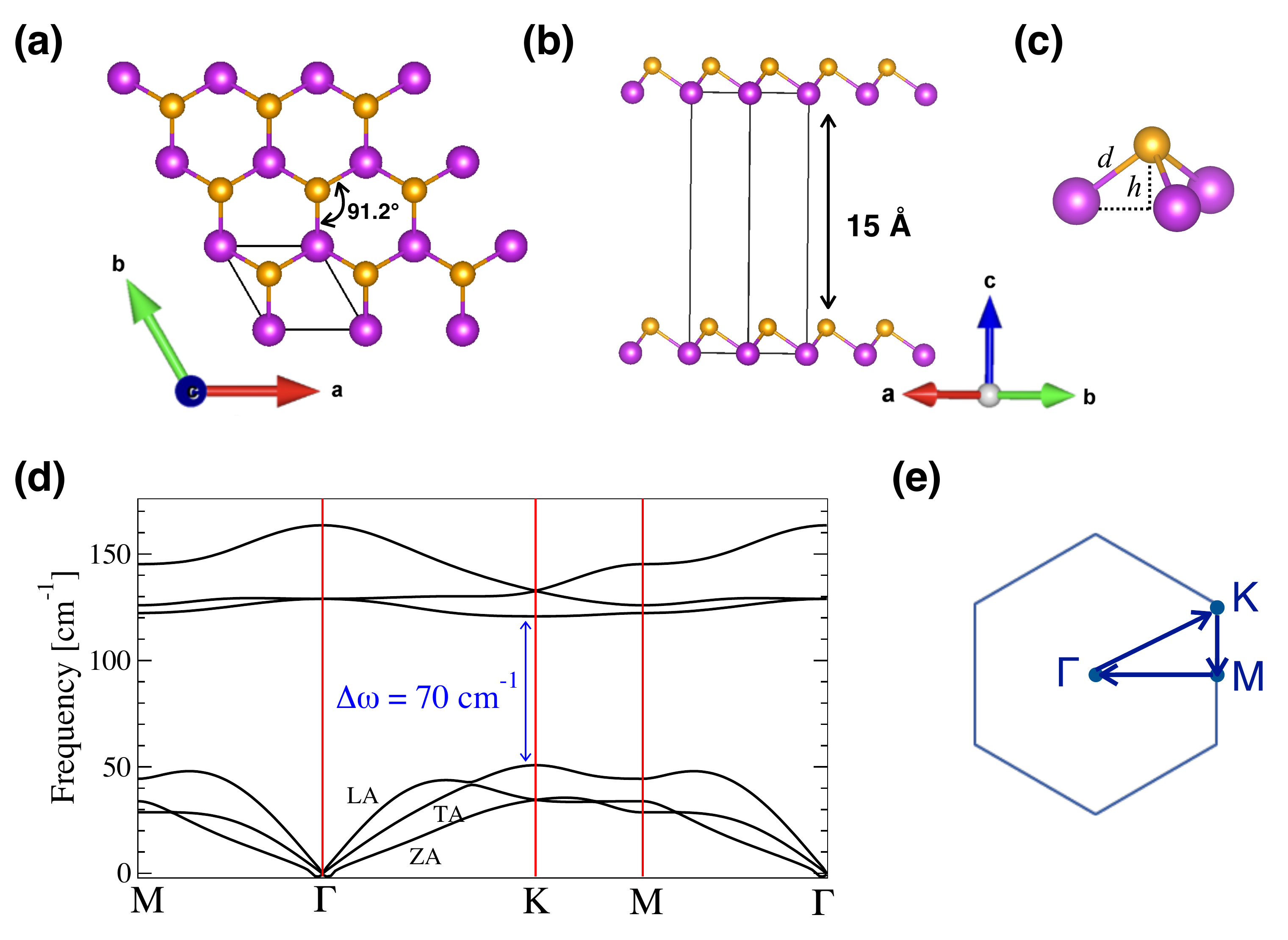}
 \caption{(Color online) Figure (a-b) represent the honeycomb crystal structure of BiSb monolayer viewed from top and side directions. Bi atoms are shown in purple color while Sb atoms are shown in orange color. (c) Bonding of Bi-Sb atoms. Here, $d$ represents the Bi-Sb bond length and $h$ represents the buckling height. (d) The phonon spectra of BiSb monolayer calculated along the high symmetry directions of Brillouin zone. (e) The hexagonal 2D Brillouin zone labelled with high symmetry points. 
 \label{fig:mono_crystal-phonons}}
 \end{figure}

Fig.~\ref{fig:mono_crystal-phonons}(a-c) shows the crystal structure of an isolated BiSb monolayer. The fully optimized structure exhibits a buckled honeycomb lattice having a three fold rotational symmetry in $P3m1$ space group (156). The optimized lattice parameters are $a = b = 4.255$ \AA. The Bi-Sb bond length ($d$) and the buckling height ($h$) are 2.98 and 1.69 \AA, respectively. The Bi-Sb-Bi bond angle is 91.2\degree. In order to test the energetic stability of BiSb monolayers, we calculate the formation energy of a single optimized BiSb monolayer using the following expression: 

\begin{equation}
  {E{_{formation}} =\frac{E_{coh}(BiSb) - E_{coh}(Bi) - E_{coh}(Sb)}{N_{atom}}}\label{eqn:1}
\end{equation}

where, $E_{coh}$ denotes the cohesive energy relative to the free constituent atom and $N_{atom}$ is the total number of atoms in the cell. The calculated formation energy ($E_{formation}$) of a single BiSb monolayer is $-2.23$ eV/atom, which is lower than the formation energy of most of the previously synthesized single layer transition-metal dichalcogenides.\cite{Coleman_Science2011}  For example, the formation energy of single layer MoS$_2$, MoSe$_2$, MoTe$_2$, WS$_2$, WSe$_2$, WTe$_2$, NbTe$_2$ and NbSe$_2$ is $-0.83, -0.75, -0.30, -0.91, -0.62, -0.10, -0.42$ and $-0.88$ eV/atom, respectively.\cite{Ataca_JPCC2012} However, the formation energy of BiSb monolayer is slightly larger than that of other binary V-V monolayer compounds such as PN, AsN, SbN, AsP, SbP and SbAs.\cite{yu2016atomically} Nonetheless, BiSb monolayer forms an energetically stable crystal structure and therefore, it can be synthesized in laboratory. 

We test the dynamical stability of the BiSb monolayer by analyzing the phonon spectra. Fig.~\ref{fig:mono_crystal-phonons}(d) represents the phonon bandstructure calculated along the high symmetry directions of Brillouin zone shown in Fig.~\ref{fig:mono_crystal-phonons}(e). We notice that there are three optical and three acoustical phonon branches, corresponding to total 6 branches due to 2 atoms per cell. The longitudinal acoustic (LA) and transverse acoustic (TA) modes represent in-plane vibrations while the ZA branch corresponds to the out-of-plane vibrations. One can notice that all phonon frequencies are positive, which confirms the dynamical stability of BiSb monolayer at 0 K. It is worthy to note that the ZA phonon branch exhibits a small imaginary frequency having a `U'-shape near the $\Gamma$-point. This `U'-shape feature does not correspond to the lattice instability, nevertheless, it is a signature of the flexural acoustic mode present in two-dimensional systems.\cite{jiang2015review} The flexural acoustic modes are known to play a crucial role in governing the thermal and mechanical properties of two-dimensional systems.  Such `U'-shape feature near the $\Gamma$-point has been observed in many other similar two-dimensional systems studied earlier.\cite{ZhengPRB2015, yu2016atomically, wang2016strain} Thus, our phonon calculations confirm that BiSb monolayers are dynamically stable and therefore, it can exist as a free standing two-dimensional crystals. Furthermore, we notice that the acoustic and optical phonon branches are well separated, indicating good optical response of BiSb monolayer. There is a wide direct frequency gap ($\Delta$$\omega$) of 70 cm$^{-1}$ between the optical and acoustic phonon modes at K-point. In the photoexcitation experiments and in the solar cell applications, the excitons (photon excited electron-hole pairs) loose most of their energy by exciting the optical phonons. These excited optical phonons further decay into the acoustic phonons which carry the heat away. Notably, in our case the phonon frequency gap is larger than the frequency of the hardest acoustic phonon mode (\~50 cm$^{-1}$). This feature significantly avoids the Klemens decay,\cite{KlemensPRB1966} which indicates the possibility to use such type of materials for the high-efficiency solar cell applications. \cite{konig2010hot} 

From the application point of view, it is important to check the mechanical stability of any new material. Therefore, we have performed first-principle calculations to determine the elastic constants (C$_{ij}$) of BiSb monolayer. The elastic constants were calculated using the stress-strain relations as implemented in the {\sc VASP} code \cite{Kresse1996, Kresse1999}. We used a dense $21\cross21\cross1$ $\Gamma$-kmesh to reduce the numerical errors caused by the in-plane strain of the system. The second Piola--Kirchhoff stress method was used to determine the elastic constants in 2D with units of N/m.\cite{QPeng2013_ZnO, QPeng2013} We find that $C_{11}$ = 24.4 N/m and $C_{12}$ = 5.8 N/m. We  further calculated the 2D layer modulus ($\gamma$), Young's modulus (${Y^{2D}_S}$) and Poisson's ratio ($\nu$) using the method suggested by Andrew et al.\cite{AndrewPRB_2012} The obtained values of $\gamma$, ${Y^{2D}_S}$ and $\nu$ are 15.12 N/m, 23.0 N/m and 0.24, respectively. Poisson's ratio is larger compared to that of h-BN (0.22) and graphene (0.17) monolayers, however, $\nu$ is smaller compared to that of buckled Si (0.34) and Ge (0.28) monolayers.\cite{AndrewPRB_2012} $\gamma$ and ${Y^{2D}_S}$ values of BiSb monolayer are significantly lower than that of h-BN and graphene. This is due to the fact that Bi-Sb bonds are weaker compared to B-N and C-C bonds.  All the obtained elastic constants are positive confirming the mechanically stability of BiSb monolayer. A detailed study of the mechanical properties of strained BiSb monolayer will be published elsewhere. 

\begin{figure}[htb!]
 \centering
 \includegraphics[width=17.0cm, keepaspectratio=true]{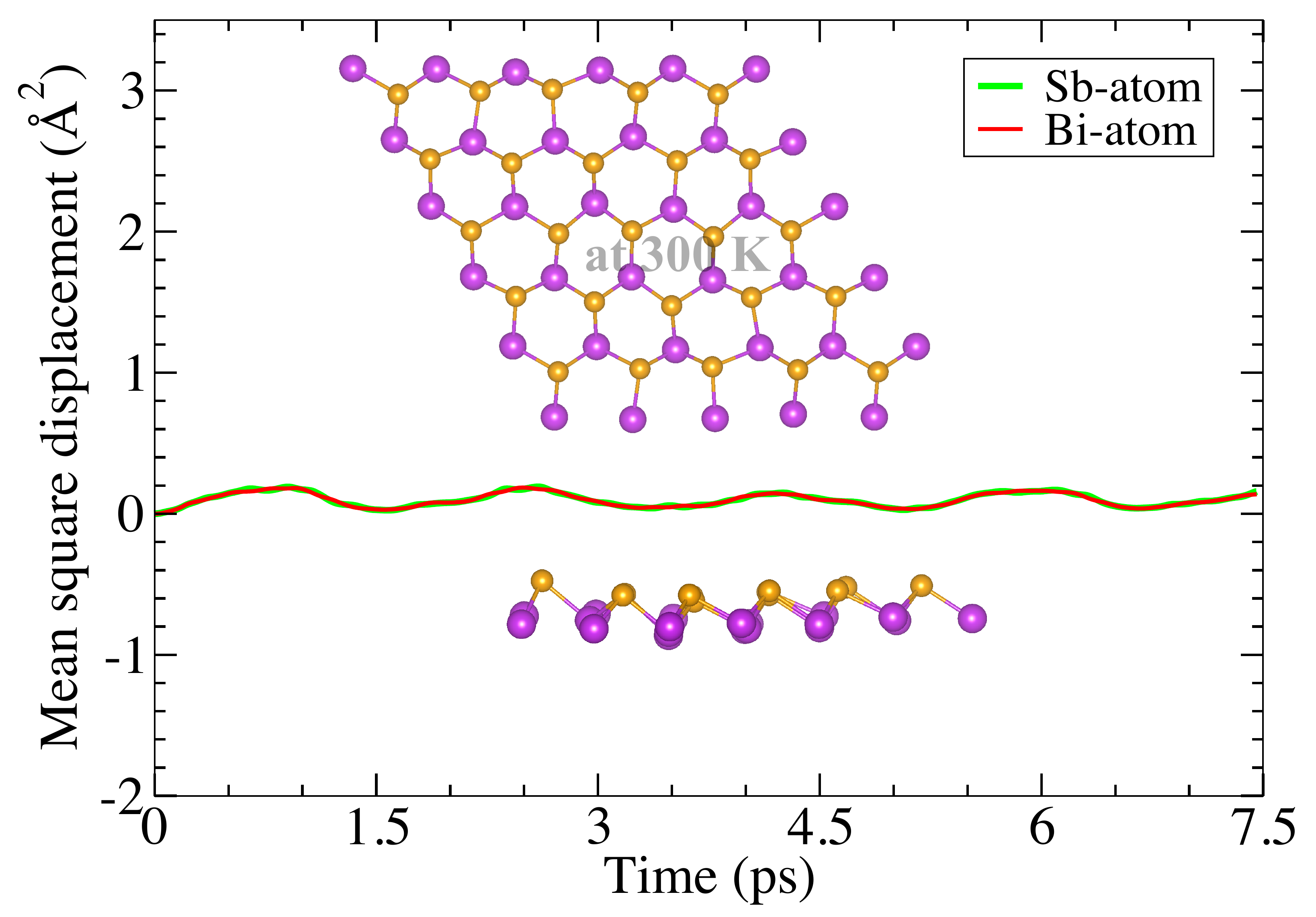}
 \caption{(Color online) This figure shows variation in the mean-square displacement of Bi (purple) and Sb (orange) atoms as a function of simulation time at 300~K. Insets show the top and side views of the geometric structures of monolayer at the end of MD simulation at 300~K. 
 \label{fig:md_300K}}
 \end{figure}

We further perform the \emph{ab-initio} MD simulations to assess the thermal stability of BiSb monolayer at room temperature. Fig.~\ref{fig:md_300K} shows the mean-square displacement of the Bi and Sb atoms as a function of the simulation time (in ps). The snapshots of the geometric structure at 300~K obtained at the end of the MD simulation (total time =  7.5 ps) are also given in the insets of Fig.~\ref{fig:md_300K}. We see that the mean-square displacement oscillates with an amplitude of about 2\% of the total Bi-Sb bond length. Additionally, one can notice that the geometric structures at 0~K (Fig.~\ref{fig:mono_crystal-phonons}) and at 300~K (insets of Fig.~\ref{fig:md_300K}) are quite similar. This confirms the dynamical stability of BiSb monolayer at the room temperature, suggesting the possible applicability of this system for room temperature device applications. 


\subsection{Electronic structure: the giant 2D Rashba spin-splitting}
\begin{figure}[hb!]
 \centering
 \includegraphics[width=17.0cm, keepaspectratio=true]{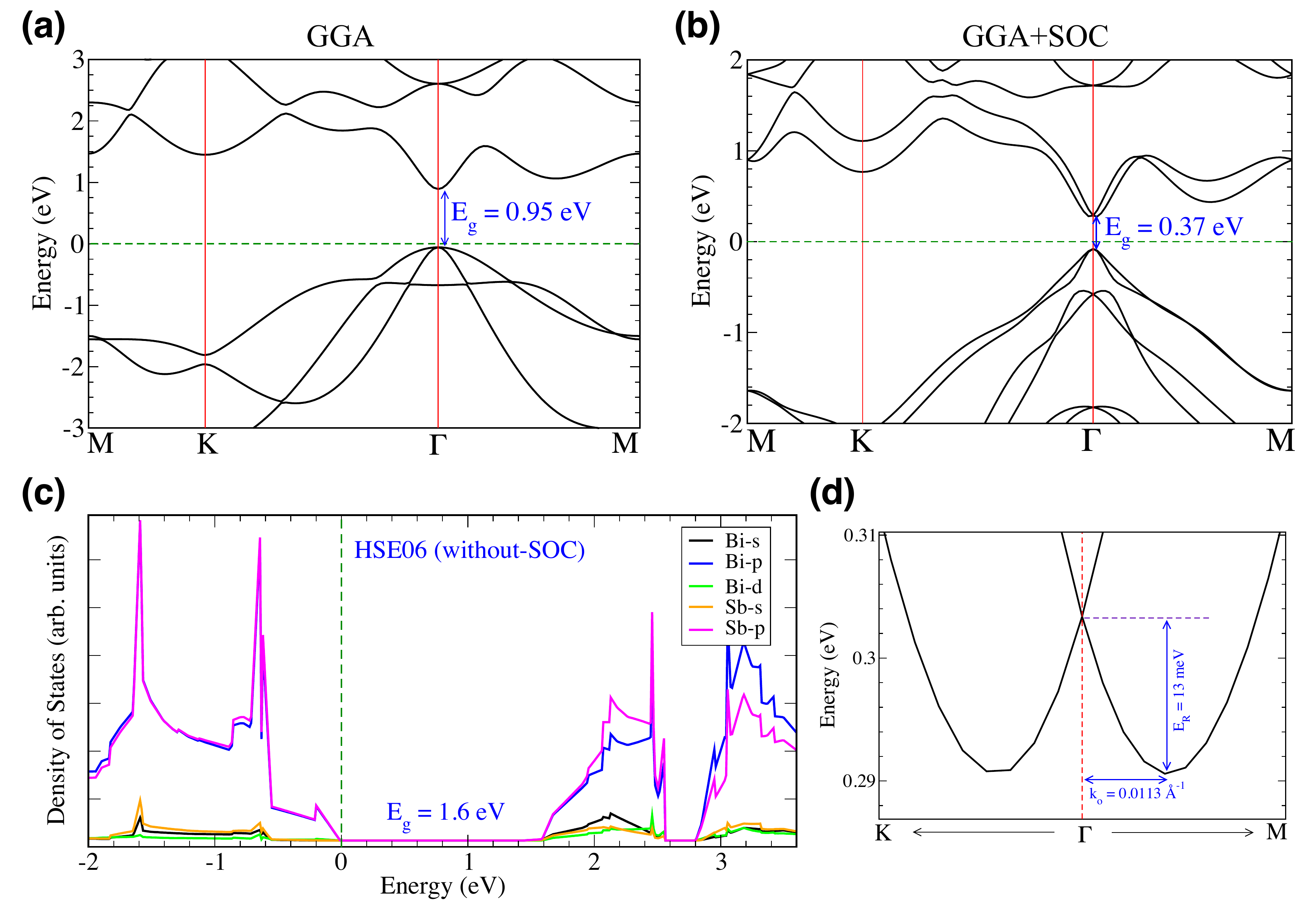}
 \caption{(Color online) The electronic bandstructure of BiSb monolayer, (a) without-SOC (b) with-SOC. (c) The atomic orbital projected DOS calculated using HSE06 approximation and a $\Gamma$ k-mesh of size $21\cross21\cross1$. (d) The enlarged view of Rashba spin-splitting of conduction bands near the Fermi-level. The green dotted line represents Fermi-level.
 \label{fig:bands_dos}}
 \end{figure}

Fig.~\ref{fig:bands_dos} shows the electronic bandstructure of BiSb monolayer calculated using (a) GGA and (b) GGA+SOC approximations. Without-SOC, the bandstructure reveals the semiconducting nature with presence of a direct bandgap ($E_g$) of 0.95 eV at the $\Gamma$ point. However, bulk BiSb is known to exhibit an indirect bandgap semiconducting nature \cite{sobhit2016PRB}. The thickness dependent transition from an indirect (3D) to direct (2D) bandgap semiconductor is not new in layer-like systems, and it has already been reported for other layer-like systems.\cite{Mak_PRL2010} In BiSb monolayer, the conduction band bottom (CBB) shows a parabolic nature near $\Gamma$ point indicating the presence of highly mobile light electrons (nearly free-electrons), while the valence band top (VBT) indicates the presence of relatively heavy holes near $\Gamma$ point. Since DFT is known to underestimate the energy bandgap, we employ the Heyd-Scuseria-Ernzerhof (HSE06) hybrid functional to predict a more reasonable energy bandgap. Fig.~\ref{fig:bands_dos}(c) shows the orbital projected DOS calculated using HSE06 approximation (without-SOC) for BiSb monolayer. We find that the HSE06 bandgap is approximately 1.6 eV which is larger than the DFT (GGA and GGA+SOC) predicted bandgap and lies in the visible range of electromagnetic spectrum. This makes this system of peculiar interest for optoelectronic applications. It is worth mentioning that the HSE06 functional only enhances the bandgap without changing the shape of the electronic bands. Our calculations further indicate that the bands near Fermi-level (E$_F$) are mainly composed of Bi $6p$ and Sb $5p$ orbitals. 

Inclusion of SOC results in the spin-splitting of the electronic bands (Fig.~\ref{fig:bands_dos}(b)). Remarkably, we notice a Rashba type spin-splitting of CBB near $\Gamma$-point, while the VBT remains degenerate (Fig.~\ref{fig:bands_dos}(b, d)). Also, the direct DFT bandgap at $\Gamma$-point decreases to 0.37 eV. To quantitatively determine the strength of the Rashba spin-splitting, we calculate the Rashba parameters as described in Ref \cite{silvia_bulkgete_2013}. We find that $E_R$ = 13 meV, $k_o$ = 0.0113 \AA$^{-1}$~and $\alpha_R = 2E_R/k_o$ =  2.3 eV\AA. This is amongst the largest Rashba spin-splitting in two-dimensional materials reported so far. For example: the reported Rashba energy (Rashba constant) for Au(111) surface, InGaAs/InAlAs interface, LaAlO$_3$/SrTiO$_3$ oxide interface and Bi(111) surface is 2.1 meV (0.33 eV\AA)\cite{LaShell_AuPRL}, 0.98 meV (0.07 eV\AA)\cite{Nitta_PRL1997}, $<$5 meV (0.01--0.05 eV\AA)\cite{CavigliaPRL2010, ZhongPRB2013} and 14 meV (0.55 eV\AA)\cite{Koroteev_PRL2004}, respectively. One can notice that the Bi(111) surface exhibits the largest Rashba spin-splitting ($E_R$ = 14 meV) amongst all the above listed systems. However, the semimetallic nature of Bi(111) surface limits its applications in the spintronics devices\cite{Ast_PRL2001, Koroteev_PRL2004}. The BiSb monolayer overcomes this limitation due to its direct bandgap semiconducting nature coupled with the large Rashba spin-splitting, which is comparable with the Rashba spin-splitting of Bi(111) surface. Interestingly, we notice that the Rashba spin-splitting of BiSb monolayer is even comparable with some of the known bulk Rashba semiconductors such as BiAlO$_3$($E_R$ = 7.34--8.62 meV, $\alpha_R$ = 0.39-- 0.74\cite{PhysRevB_silveira2016}), BiTeI ($E_R$ = 100 meV, $\alpha_R$ = 3.8 eV\AA \cite{IshizakaNature2011}), BiTeCl ($E_R$ = 18.5 meV, $\alpha_R$ = 1.2 eV\AA \cite{XiangPRBbitecl_2015}), GeTe ($E_R$ = 227 meV, $\alpha_R$ = 4.8 eV\AA \cite{silvia_bulkgete_2013}), SnTe ($E_R$ = 272 meV, $\alpha_R$ = 6.8 eV\AA \cite{SnTe_PRBsilvia2014}), LiZnSb ($E_R$ = 21 meV, $\alpha_R$ = 1.82 eV\AA \cite{NarayanPRB2015}) and KMgSb ($E_R$ = 10 meV, $\alpha_R$ = 0.83 eV\AA \cite{NarayanPRB2015}). Thus, after comparing the strength of the Rashba spin-splitting in BiSb monolayer and several other known large Rashba semiconductors, one can realize that the Rashba spin-splitting of the highly mobile light-electrons in BiSb monolayer is indeed remarkable and therefore, this system can be used to make efficient advanced spin-field effect transistors. 

\begin{figure}[hb!]
 \centering
 \includegraphics[width=18cm, keepaspectratio=true]{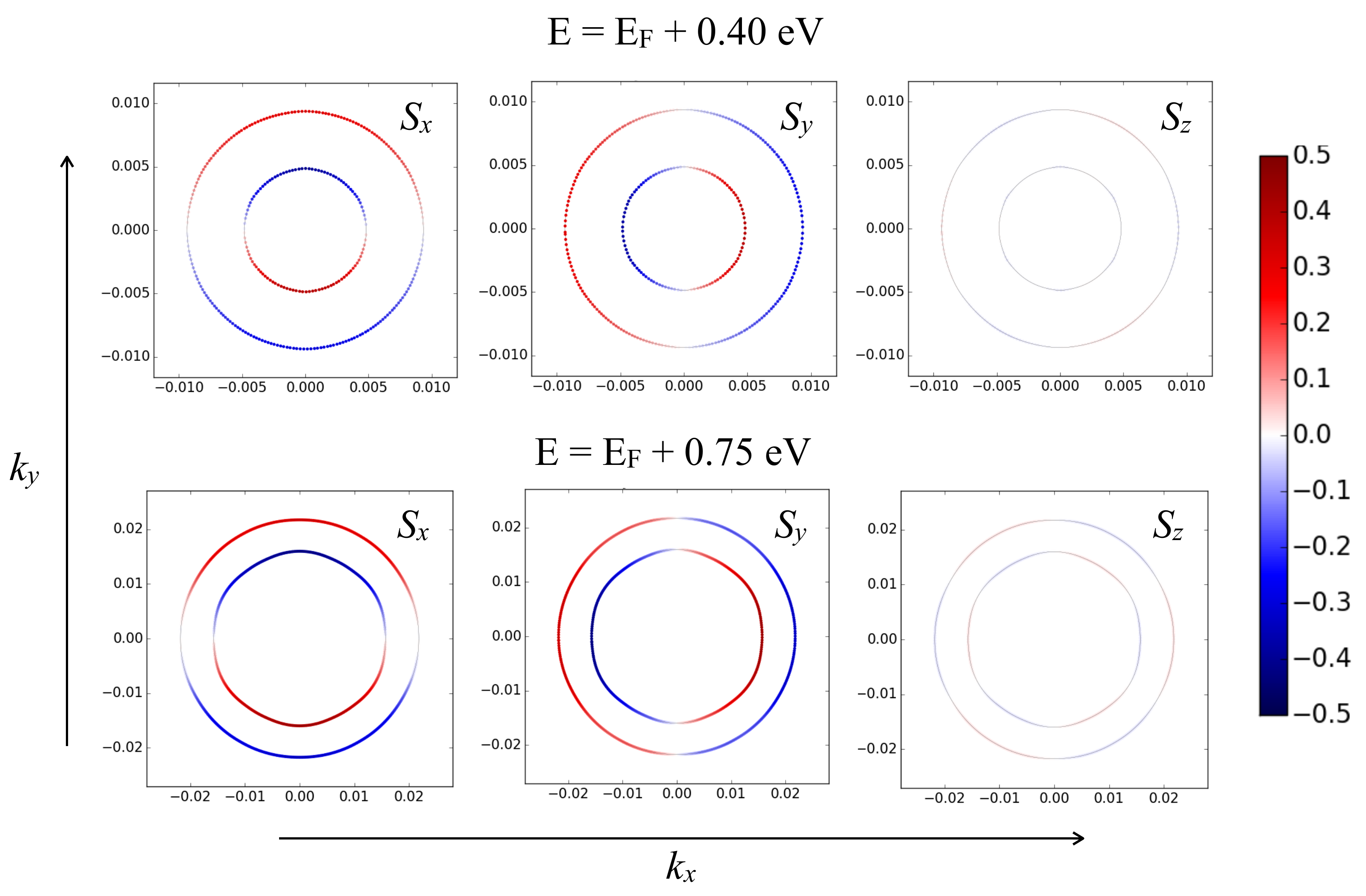}
 \caption{(Color online) Spin projected constant energy contour plots of spin-texture calculated in a $k_x-k_y$ plane centered at the $\Gamma$-point. Top (Bottom) row represents the spin-textures calculated at an energy surface 0.40 eV (0.75 eV) above the Fermi-level. In the color scale, red color depicts spin-up states while blue color depicts spin-down states. 
 \label{fig:spin_texture}}
 \end{figure}

Fig.~\ref{fig:spin_texture} shows the constant energy 2D contour plots of spin-texture calculated in a $k_x-k_y$ plane centered at the $\Gamma$-point. Evidently, one can notice the Rashba type spin-splitting of spin-up (red) and spin-down (blue) electronic bands. The concentric spin-texture circles are result of the purely two-dimensional Rashba spin-splitting present above the Fermi-level (conduction bands). A further analysis of the projection of different spin-components ($S_x$, $S_y$ and $S_z$) on the electronic bands reveals that only in-plane $S_x$ and $S_y$ spin components are present in the Rashba-spin split bands, without presence of any out-of-plane $S_z$ component.  This further confirms that the spin-splitting of electrons in BiSb monolayer has purely two-dimensional Rashba nature. Interestingly, we observe the Rashba spin-splitting at significantly large energies (0.40 and 0.75 eV) above the Fermi-level. 

\subsection{Effect of in-plane bi-axial strain on the electronic properties}
\begin{figure}[hb!]
 \centering
 \includegraphics[width=15cm, keepaspectratio=true]{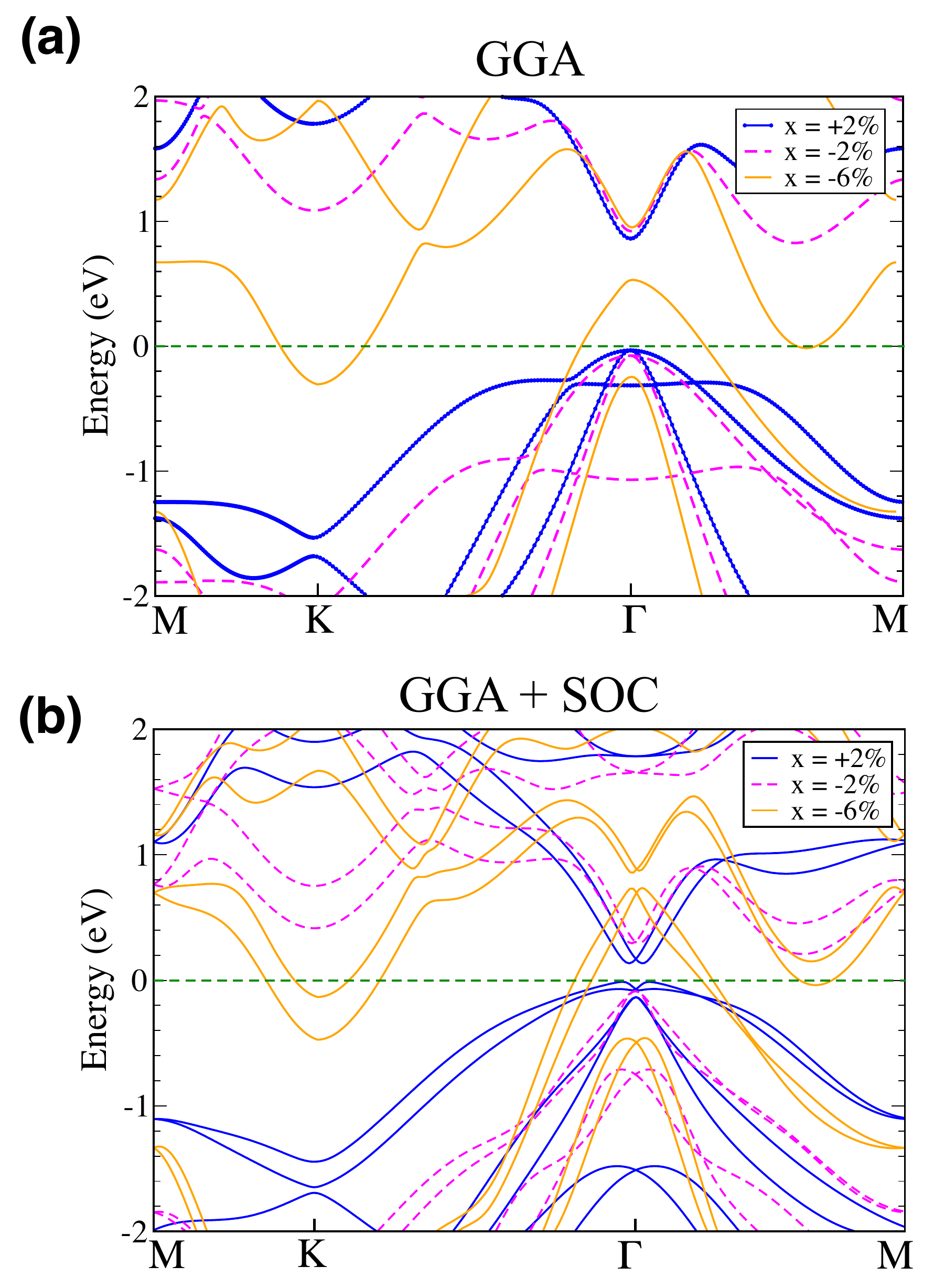}
 \caption{(Color online) The electronic bandstructure of strained BiSb monolayers at an applied in-plane bi-axial strain $x = +2\%$ (stretched), $-2\%$ and $-6\%$ (compressed) monolayers, (a) without-SOC and (b) with-SOC. The green dotted line represents Fermi-level.
  \label{fig:biaxial_strain}}
 \end{figure}

It is very important to study the effect of the bi-axial strain on the electronic properties before we shift our concentration to the BiSb/AlN heterostructure systems.  Therefore, we study the strained BiSb monolayers by applying bi-axial strain ($x$) ranging from $-8\%$ (compression) to $+8\%$ (elongation) on the $a$ and $b$ lattice vectors. Fig.~\ref{fig:biaxial_strain} shows change in the with- and without-SOC electronic bandstructures with respect to the $x$ = $+2\%$ (solid blue lines), $-2\%$ (broken magenta line) and $-6\%$ (solid orange lines) bi-axial strains. From Fig.~\ref{fig:biaxial_strain}(a-b), we notice that the system retains its direct bandgap semiconducting nature until $x = -2\%$. However, with further increase in bi-axial strain, it changes to an indirect bandgap semiconductor at $-4\%$ and eventually, we realize a metallic phase-transition at $x = -6\%$. Furthermore, the degeneracy of the valence band near $\Gamma$-point breaks down at $x$ = $-6\%$ bi-axial compression and a Rashba type spin-splitting appears in the valence band states. Analysis of the Rashba spin-splitting in the strained bands evinces that the Rashba energy (constant) decreases from 13 meV (2.3 eV\AA) to 9.21 meV (1.77 eV\AA) with increase in the bi-axial compression from $x = 0\%$ to $x = -4\%$. However, we observe significant increase in the Rashba spin-splitting parameters with increasing bi-axial elongation (For $x = +6\%$ elongation: $E_R$ = 33.4 meV, $\alpha_R$ = 3.56 eV\AA). Consequently, one can significantly tune the strength of the Rashba spin-splitting by varying the in-plane bi-axial strain. Nonetheless, in order to experimentally realize the novel direct bandgap Rashba semiconducting properties of the BiSb monolayer, we must grow this system on top of a substrate such that the lattice mismatch is less than $2\%$. This can be experimentally realized by a proper choice of substrate material.

\subsection{BiSb/AlN heterostructures}
\begin{figure}[hb!]
 \centering
 \includegraphics[width=18cm, keepaspectratio=true]{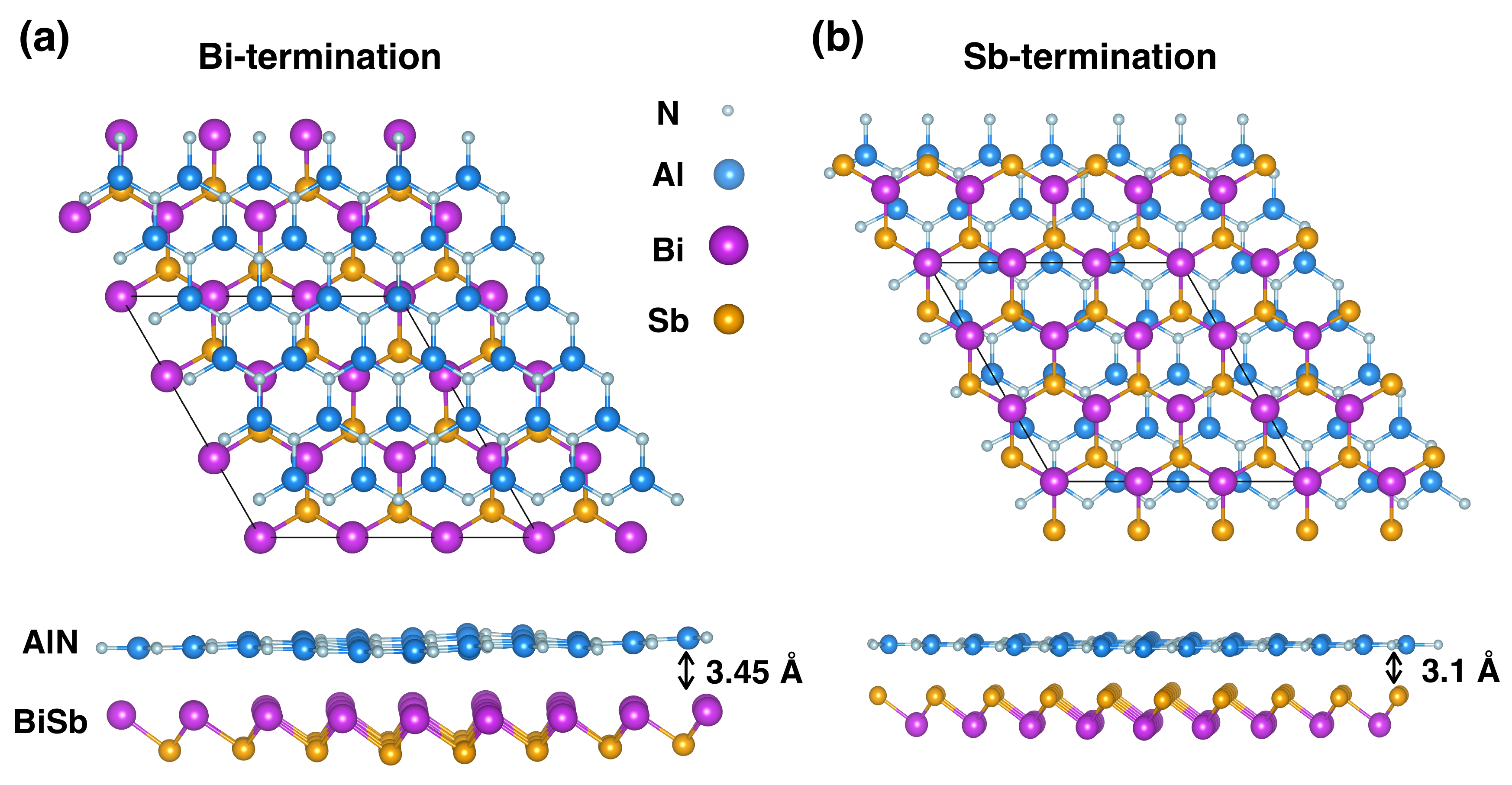}
 \caption{(Color online) Figure represents the relaxed geometric structures of BiSb/AlN heterostructures with two possible terminations at the interface. Fig. (a) shows the top and side views of Bi-terminated interface while Fig. (b) shows the top and side views of Sb-terminated interface. 
 \label{fig:heter_crystal}}
 \end{figure}

How do the electronic properties of BiSb monolayer change in presence of another substrate material? Can we still realize the Rashba spin-splitting of 2DEG present at the interface of heterostructure systems? How would the energetic stability of an isolated BiSb monolayer change in presence of another contact material? In order to address these questions, we model BiSb/AlN heterostructures and study their electronic properties. To minimize the lattice mismatch between two stackings, we build the heterostructure using a $4\cross4\cross1$ supercell of AlN and a $3\cross3\cross1$ supercell of BiSb. Thus obtained lattice mismatch is 1.3\%. One could also use hexagonal BN or GaN as a substrate material for BiSb monolayer. The lattice mismatch for BiSb($3\cross3\cross1$)/BN($5\cross5\cross1$) and BiSb($3\cross3\cross1$)/GaN($4\cross4\cross1$) combinations is 1.7\% and 2.15\%, respectively. In the present work, we study the BiSb/AlN heterostructure system because this system exhibits minimum lattice mismatch. Fig.~\ref{fig:heter_crystal} shows the BiSb/AlN heterostructures with two possible layer terminations at the interface: first, Bi-termination (Fig.~\ref{fig:heter_crystal}(a)) and second Sb-termination (Fig.~\ref{fig:heter_crystal}(b)). The vdW optimized lattice constants for Bi- and Sb-terminated interfaces are 12.541 and 12.536 \AA, respectively. In both cases, the BiSb lattice is being compressed by 1.8\% while the AlN lattice is being stretched by 0.5\%. The interlayer spacing between AlN and BiSb monolayers is 3.45 \AA~(3.10 \AA) for Bi-terminated (Sb-terminated) interfaces. The presence of larger interlayer spacing for Bi-terminated interface can be ascribed to the fact that Bi-atom has larger covalent radius compared to the Sb-atom. 

The calculated formation energies for Bi- and Sb-terminated interfaces are $-5.94$ and $-5.95$ eV/atom, respectively. This suggests that the Sb-terminated interface is energetically 10 meV/atom more favorable compared to the Bi-terminated interface. Furthermore, comparing the formation energy of an isolated BiSb monolayer with that of BiSb/AlN system, we notice that the BiSb/AlN heterostructures are energetically more stable than a single sheet of BiSb. One can switch the ferroelectric polarization, and thus the electric field direction at the interface, by switching the Bi and Sb layers. An inverted ferroelectric polarization leads to the complete inversion of the spin-polarized Rashba bands near the Fermi-level. This particular feature has already been investigated for bulk BiSb. \cite{sobhit2016PRB} We observe that BiSb monolayer exhibits the same behavior upon reversal of the ferroelectric polarization. 

\begin{figure*}[htb!]
 \centering
 \includegraphics[width=18.5cm, keepaspectratio=true]{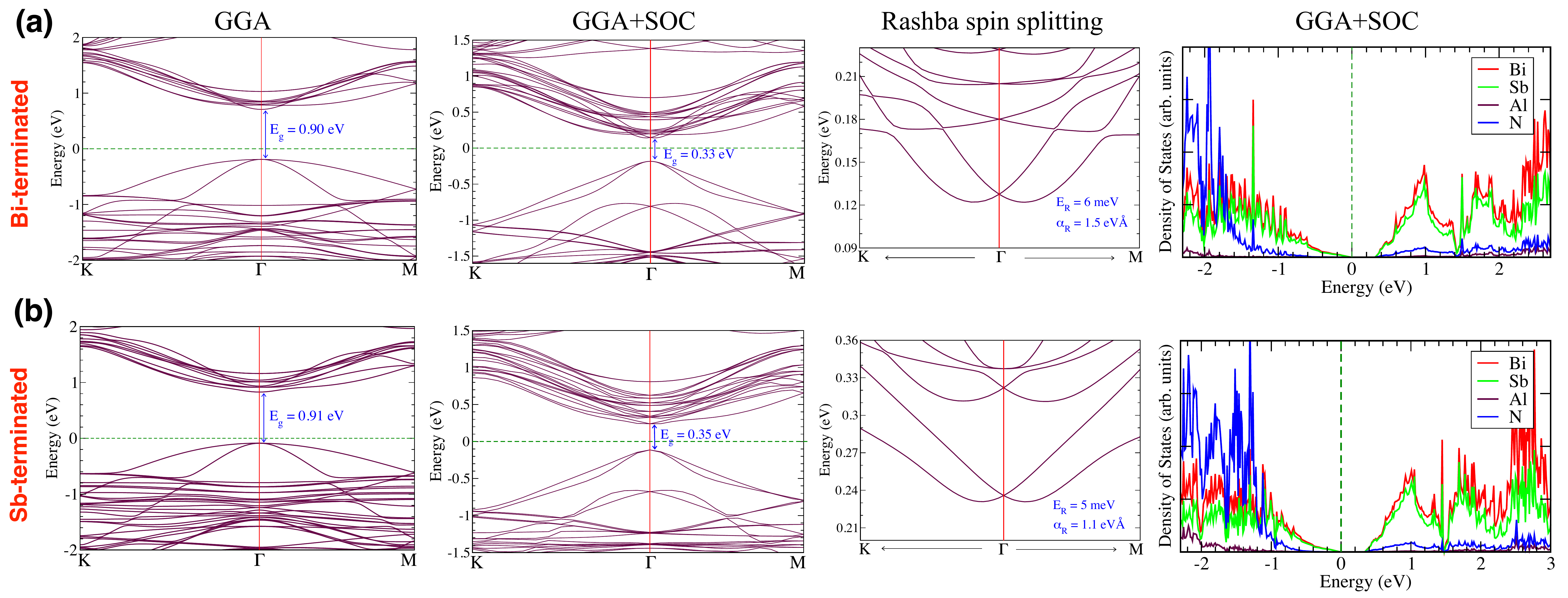}
 \caption{(Color online) Top (Bottom) row of figure shows the electronic bandstructure calculated without-SOC, with-SOC, enlarged conduction bands showing Rashba splitting near Fermi-level, and the projected DOS calculated using a $\Gamma$ k-mesh of size $11\cross11\cross1$ for Bi-terminated (Sb-terminated) interface. The vdW interaction was included in all first-principle calculations for BiSb/AlN heterostructures. The green dotted represents Fermi-level.
 \label{fig:hetero_bands_dos}}
\end{figure*}

The calculated electronic bandstructure and projected DOS of BiSb/AlN heterostructures are shown in Fig.~\ref{fig:hetero_bands_dos}. The top row (Fig.~\ref{fig:hetero_bands_dos}(a)) shows the electronic bands without-SOC,  with-SOC, conduction bands near $\Gamma$-point showing Rashba spin-splitting and the projected DOS for Bi-terminated interface, while the bottom row (Fig.~\ref{fig:hetero_bands_dos}(b)) shows the same but for Sb-terminated interface. Without-SOC, both Bi- and Sb-terminated heterostructures show a direct bandgap of 0.90 and 0.91 eV at the $\Gamma$-point, respectively. The CBB of both heterostructures has a free-electron gas like parabolic shape near the $\Gamma$-point whereas the top valence bands are quite flat in both heterostructures indicating presence of heavy holes along with existence of light 2DEG at the interface. The direct gap of the heterostructures is lower compared to that of the isolated BiSb-monolayer. Interestingly, we notice a Rashba type of spin-splitting of CBB in the SOC bands of both heterostructures. Also, the direct DFT gap of both heterostructures reduces to 0.33 eV (Bi-terminated) and 0.35 eV (Sb-terminated) due to inclusion of SOC. The calculated strength of the Rashba spin-splitting is $E_{R}$ = 6 meV and $\alpha_{R}$ = 1.5 eV\AA~ for Bi-terminated interface and $E_{R}$ = 5 meV and $\alpha_{R}$ = 1.1 eV\AA~for Sb-terminated interface. The strength of the Rashba splitting is larger at the Bi-terminated interface. This is due to the fact that Bi has larger SOC compared to the Sb atom \cite{BiSOC_PRB2007, BiSOC_PRL2007, SbSOC_PRB2008}. It is noteworthy that the strength of the Rashba spin-splitting is amongst the largest yet known Rashba heterostructure systems.  

\begin{figure}[htb!]
 \centering
 \includegraphics[width=18cm, keepaspectratio=true]{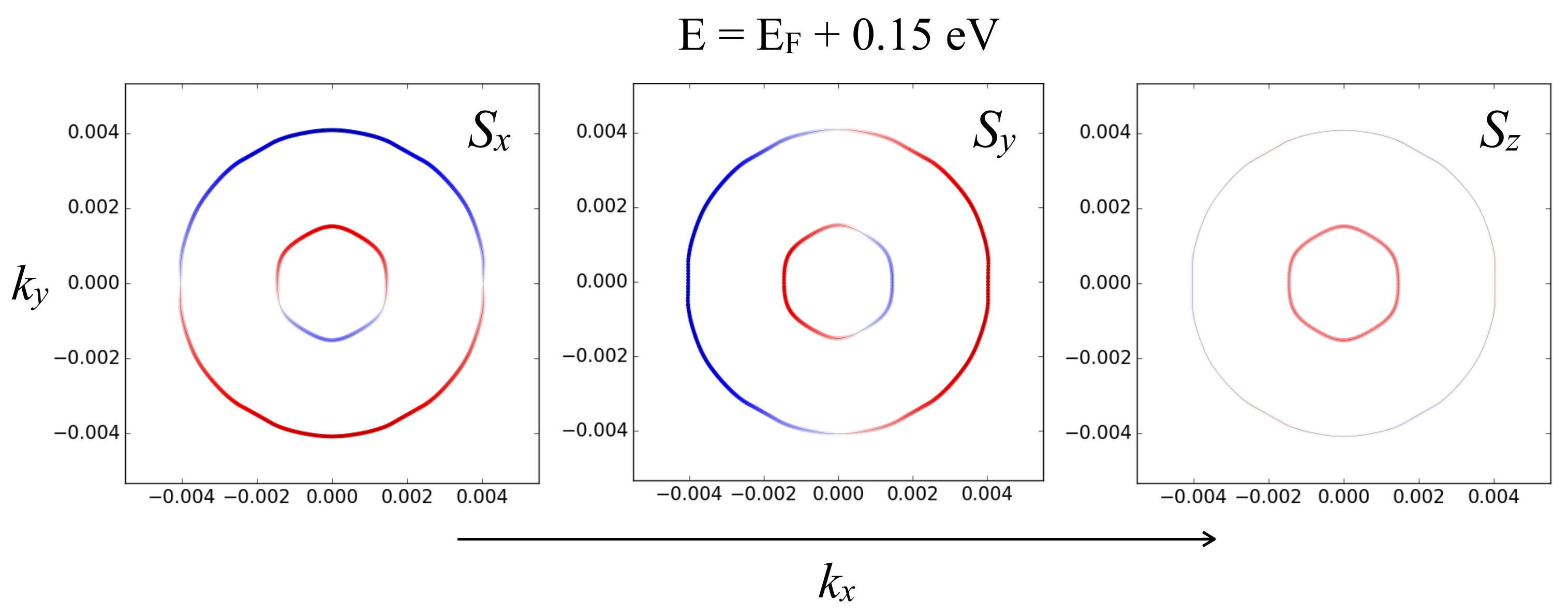}
 \caption{(Color online) Spin-texture for Bi-terminated interface calculated in a $k_x-k_y$ plane centered at the $\Gamma$ point and at an energy surface 0.15 eV above the Fermi-level. The red color depicts spin-up states while blue color depicts spin-down states. 
 \label{fig:hetero_spin_texture}}
\end{figure}

To further confirm the 2D nature of Rashba spin-splitting, we calculate the spin projected spin-texture in a $k_{x}$-$k_{y}$ plane centered at the $\Gamma$-point. The constant energy (0.15 eV above Fermi-level) contour plots of the spin-texture for Bi-terminated interface are shown in Fig.~\ref{fig:hetero_spin_texture}. The spin-texture for Sb-terminated interface exhibits similar shape. The circular shape of the spin-textures confirm the purely 2D nature of Rashba spin-splitting of 2DEG at the interface. Furthermore, we observe that only $S_x$ and $S_y$ spin-components are present in $k_{x}$-$k_{y}$ plane without presence of any $S_z$ component. This indicates that the spin-texture has only in-plane spin-components. These remarkable giant Rashba features make this system suitable for many interesting applications in the spintronics industry. In particular, one can make highly efficient spin field-effect transistors using this system, where we exploit the giant 2D Rashba effect to control the spin-state of 2DEG. 

In summary, we have identified a new energetically, dynamical and mechanical stable crystal phase of BiSb monolayer which has buckled honeycomb lattice geometry. Our first principle calculations reveal that this monolayer is a direct bandgap semiconductor having free-electron-like parabolic conduction band features near $\Gamma$-point. Inclusion of SOC yields Rashba type spin-splitting of conduction bands near Fermi-level. The obtained Rashba spin-splitting parameters ($E_R$ = 13 meV, $\alpha$$_R$ = 2.3 eV\AA) suggest that this system is amongst the largest yet known 2D Rashba semiconductors. The constant energy contour plots of spin-textures confirm the purely two-dimensional nature of the Rashba splitting. Both BiSb/AlN heterostructures exhibit direct bandgap semiconducting nature having parabolic conduction bands near Fermi-level, which indicates the presence of 2DEG at the interface. The strength of the Rashba splitting is stronger for the Bi-termination ($E_R$ = 6 meV, $\alpha$$_R$ = 1.5 eV\AA) compared to the Sb-terminated interface ($E_R$ = 5 meV, $\alpha$$_R$ = 1.1 eV\AA). The presence of direct bandgap with light electron-like features makes this system interesting for 2D optical device applications. Additionally, the existence of two-dimensional giant Rashba effect together with the electronic bandgap in visible range paves the way to use BiSb-monolayer and BiSb/AlN heterostructure systems in the development of highly efficient spin field-effect transistors and optoelectronics devices in spintronics industry. 

\begin{acknowledgement}
This work used the Extreme Science and Engineering Discovery Environment (XSEDE), which is supported by National Science Foundation grant number OCI-1053575. Additionally, the authors acknowledge the support from Texas Advances Computer Center (TACC), Bridges supercomputer at Pittsburgh Supercomputer Center and Super Computing Systems (Spruce and Mountaineer) at West Virginia University (WVU). A. H. R. and S. S. acknowledge the support from National Science Foundation (NSF) DMREF-NSF 1434897 and DOE DE-SC0016176 projects. S. S. thanks the donors of Jefimenko family for their financial support through the Oleg D. and Valentina P. Jefimenko Physics Fellowship at WVU. 
\end{acknowledgement}

\bibliography{achemso-BiSb}

\newpage
\section{Figures}

\end{document}